\documentclass[pre,epsfig,twocolumn,showpacs,preprintnumbers,amssymb,eqsecnum]{revtex4}
\usepackage{bm,graphicx,amsmath}

\newcommand{\calI}{{\mathcal I}}

\begin{document}

\title{Poisson-Bracket Approach to the Dynamics of Nematic Liquid
     Crystals. The Role of Spin Angular Momentum}

\author{H.~Stark$^{1}$}
\author{T.C. Lubensky$^{2}$}
\affiliation{$^1$Universit\"at Konstanz, Fachbereich Physik, D-78457
             Konstanz, Germany\\
             $^{2}$Department of Physics and Astronomy, University of
             Pennsylvania, Philadelphia PA 19104, USA}

\begin{abstract}
Nematic liquid crystals are well modeled as a fluid of rigid rods.
Starting from this model, we use a Poisson-bracket formalism to
derive the equations governing the dynamics of nematic liquid
crystals. We treat the spin angular momentum density arising from
the rotation of constituent molecules about their centers of mass
as an independent field and derive equations for it, the mass
density, the momentum density, and the nematic director. Our
equations reduce to the original Leslie-Ericksen equations,
including the inertial director term that is neglected in the
hydrodynamic limit, only when the moment of inertia for angular
momentum parallel to the director vanishes and when a dissipative
coefficient favoring locking of the angular frequencies of
director rotation and spin angular momentum diverges. Our
equations reduce to the equations of nematohydrodynamics in the
hydrodynamic limit but with dissipative coefficients that depend
on the coefficient that must diverge to produce the
Leslie-Ericksen equations.
\end{abstract}


\maketitle

\section{Introduction} \label{sec:Intro}
The Leslie-Ericksen (LE) equations \cite{LE} for the dynamics of
nematic liquid crystals have been a bulwark of liquid crystal
science since they were first derived over forty years ago.
For a historical account of their derivation see Ref.\ \cite{Carlsson99}.
They indisputably provide a correct theoretical description of the
almost limitless variety of dynamical phenomena that nematic
liquid crystals can exhibit, from simple shear flow to
hydrodynamic instabilities to complex switching in display cells.

The equations originally derived by Leslie and Ericksen are not
completely hydrodynamical.  They contain an inertial term in the
equation for the director, $\bm{n}$, specifying the direction of
molecular alignment, that leads to modes that decay in microscopic
times. When this term is ignored, the resulting equations are
purely hydrodynamical with mode frequencies that all vanish with
vanishing wavenumber.  Subsequent treatments
\cite{harvard,deGennes1993} of the dynamics of nematics produced
purely hydrodynamical equations from the outset.  To our
knowledge, all of the many experimental verifications of the
validity of the LE equations probe only the hydrodynamic limit;
they do not test the existence of or the form of the
nonhydrodynamic part of the original LE equations.

The LE equations and purely hydrodynamic treatments of
nematodynamics differ most profoundly in their treatment of
kinetic energy. In the hydrodynamic theories
\cite{harvard,deGennes1993}, the momentum density $\bm{g}$ and its
related velocity field $\bm{v}$ measure the momentum of all mass
points in the medium including those along the full length of
rigid mesogens. This momentum is a conserved quantity and is
necessarily hydrodynamic. The energy density is $g^2/(2 \rho)$,
where $\rho$ is the mass density. Since $\bm{g}$ is the total
momentum density, it contains all information about angular
momentum, and it is not necessary to introduce additional
variables to describe what we will call the spin angular momentum
associated with rotation of constituent rigid molecules about
their centers of mass\ \cite{harvard, Martin}.
The other variables that appear in the
hydrodynamic treatment are the conserved mass density and the
Frank director $\bm{n}$. In the LE treatment, there are two
contributions to the kinetic energy: a translational part $g^2/(2
\rho)$ and a rotational part, $\calI_{\perp}(\bm{n} \times
\dot{\bm{n}})^2/2 $, where $\calI_{\perp}$ is a moment of inertia
density, arising from motion of the director. The interpretation
of this decomposition \cite{LE} of the kinetic energy is that
$\bm{g}$ is now the center-of-mass momentum density, which is a
conserved variable, and that the director contribution to the
kinetic energy arises from the spin angular momentum. There are
now two contributions to the angular momentum: the spin angular
momentum and that arising from center-of-mass motion. Neither
contribution is individually conserved, but their sum is. In the
LE treatment, the equation for the director is basically an
equation for the spin angular momentum, which is neither a
conserved nor a hydrodynamical variable and which, therefore, has
nonhydrodynamic decays in it.

Though the LE equations are internally consistent and reduce to
the correct hydrodynamical form when the inertial term is ignored,
they in fact do not provide a correct description of spin angular
momentum.  Spin angular momentum is an independent dynamical
variable that is not locked to the director, though in steady
state situations it does relax to a value determined by the local
director and its rotation rate. The spin kinetic energy is
determined by the spin angular momentum and not by the dynamics of
the director. In this paper, we describe the dynamics of nematics
in terms of their non-conserved spin angular momentum density,
their conserved mass density and center-of-mass momentum density,
and their director.  For simplicity, we consider isothermal
processes only, and we ignore the equations of energy
conservation. We use the Poisson-bracket approach
\cite{Zwanzig1961,Kawasaki1970,Mori1973,Ma1975,Chaikin-Lub,Stark2003,Dzyaloshinskii1980,Grmela,Oettinger97}
to derive the equations of motion for these
variables. The hydrodynamical limit of our equations is identical
to that of the LE equations but with a slightly different
interpretation of some dissipative coefficients.  Our equations
also reduce to the full LE equations when an appropriate viscosity
diverges and spin angular momentum parallel to the director is
ignored. Previous investigations \cite{Dzyaloshinskii1980,Lubensky1970}
have noted that spin angular momentum should be treated as an
independent variable and argued that it will decay in microscopic
times to a value determined by the director. They do not, however,
provide a detailed prescription for how this decay occurs or the
conditions under which the original LE equation can be retrieved.
Finally, our equations reduce to the hydrodynamical equations for
rigid rotors on a lattice \cite{Chaikin-Lub} when coupling to
center-of-mass motion is turned off.

Since the derivation of our results is at times algebraically
tedious, we review our main results in Sec.\ \ref{sec.review}. We
present first the dynamical equations that result from our
analysis and show how they reduce to the LE equations and to true
hydrodynamical equations in the appropriate limit. In Sec.\
\ref{sec.form} we briefly summarize the Poisson-bracket formalism.
In Sec.\ \ref{sec.poisson} we introduce the fluid of rigid rods
and the relevant dynamic variables and their Poisson brackets.
Finally, in Sec.\ \ref{sec.nemato}, we derive the equations of
nematodynamics with spin angular momentum.

\section{Review of Results} \label{sec.review}

We model our nematic as a collection of uniaxial rigid rods.
The coarse-grained variables
describing this system are the mass density $\rho$, the
center-of-mass momentum density $\bm{g} = \rho \bm{v}$, the spin
angular momentum density $\bm{l}$, and the nematic director
$\bm{n}$.  The angular momentum density
$\bm{l} = \bm{\calI} \bm{\Omega}$ can be expressed in terms
of a moment-of-inertia density $\calI_{ij}= \calI_{\|}n_i n_j +
\calI_{\perp} (\delta_{ij} - n_i n_j)$ and an angular frequency
$\bm{\Omega}$. The full equations for all of these variables are
\begin{align}
&  \frac{d \rho}{d t} + \rho \bm{\nabla} \cdot \bm{v} = 0
\label{0.1} \\
& \frac{\partial g_{i}}{\partial t} = -\nabla_{j}
  \Big(\frac{g_{i}  g_{j}}{\rho }\Big)
  -\nabla_{i}p + \nabla_{j} \sigma_{ij}
\label{0.2} \\
& \frac{d \bm{n}}{d t} = \bm{\Omega} \times \bm{n} +
\frac{1}{\gamma}
  [\bm{h} + \Delta \calI \Omega_{\|}
  \bm{\Omega}_{\perp}]
\label{0.3} \\
& \calI_{\perp}
\left( \frac{d \bm{\Omega}_{\perp}}{d t} \right)_{\perp}
+ \calI_{\|} \Omega_{\|} \frac{d \bm{n}}{d t} = \bm{n} \times \bm{h}
\nonumber\\
& \qquad \qquad \qquad \quad
 -\Gamma_{\perp}^{\Omega}  (\bm{\Omega}_{\perp}   - \bm{\omega}_{\perp}) -
 \Gamma^A (\bm{A}\bm{n}) \times \bm{n} ,
\label{0.4} \\
& \calI_{\|} \frac{d \Omega_{\|}}{dt} -
\frac{\calI_{\perp}}{\gamma}(\bm{\Omega}_{\perp}\cdot \bm{h} +
\Delta \calI \Omega_{\|} \Omega_{\perp}^2) = -
\Gamma_{\|}^{\Omega} (\Omega_{\|} - \omega_{\|}) ,
\label{0.5}
\end{align}
where $\Delta \cal{I} = \cal{I}_{||} - \cal{I}_{\perp}$, $\bm{h}$
is the molecular field with components $h_i = - (\delta_{ij} - n_i
n_j) \delta F/\delta n_i$, where $F$ is the free energy, $p$ is
the pressure, $\sigma_{ij}$ is the dissipative and director part
of the stress tensor, $\bm{A}$ is the the symmetrized strain rate
tensor with components $(\partial_i v_j +
\partial_j v_i)/2$, and $\bm{\omega}= (\bm{\nabla}\times\bm{v})/2$
is half the local vorticity. In these equations, $dA/dt = \partial
A/\partial t + \bm{v}\cdot\bm{\nabla} A$ is the total derivative
of any field $A$, $c_{\|} = \bm{n} \cdot \bm{c}$ is the component
of any vector $\bm{c}$ along $\bm{n}$ and $\bm{c}_{\perp} = \bm{n}
\times (\bm{c} \times \bm{n})$ its component perpendicular to
$\bm{n}$, and $\bm{A}\bm{n}$ is the product of a matrix with a
vector with components $A_{ij} n_j$.  The stress tensor
$\sigma_{ij}$ can be decomposed into elastic, viscous symmetric,
and viscous anti-symmetric parts:
\begin{equation}
\sigma_{ij} = \sigma^{E}_{ij} + \sigma_{ij}^{S \prime} +
\sigma_{ij}^{A \prime} ,
\label{0.6}
\end{equation}
where
\begin{eqnarray}
\sigma^{E}_{ij} & = &
- \frac{\partial f}{\partial \nabla_{j}n_{k}} \nabla_{i}n_{k},\\
\sigma_{ij}^{S \prime} & = & \alpha_{ijkl} A_{kl} + \frac{1}{2}
\Gamma^{A} (\varepsilon_{ilk}n_{j}n_{l} + \varepsilon_{jlk}
n_{i}n_{l}) (\Omega_{k}-\omega_{k}) ,\nonumber \\
& & \label{0.7}\\
\sigma_{ij}^{A \prime} & = &
\frac{1}{2} \varepsilon_{ijk} \Gamma^{\Omega}_{kl}
(\Omega_{l} - \omega_{l} ) \nonumber \\
 & & \qquad + \frac{1}{2}\Gamma^{A}
(n_{j}A_{in}n_{n} - n_{i}A_{jn}n_{n}) \enspace.
\label{0.9}
\end{eqnarray}
with $\Gamma^{\Omega}_{kl} = \Gamma_{\|}^{\Omega} n_{i}n_{j}
+ \Gamma_{\perp}^{\Omega}(\delta_{ij} - n_{i}n_{j})$
and $\alpha_{ijkl}$ a fourth-rank tensor of uniaxial symmetry
given explicitly in Eq.\ (\ref{4.19}).

Equations (\ref{0.1}) to (\ref{0.9}) provide a complete
description of the dynamics of nematics. Equations (\ref{0.1}) and
(\ref{0.2}) are the familiar conservation laws for mass and
momentum.  Equation (\ref{0.3}) is the equation of motion for the
director. It is similar to that of the full hydrodynamical theory
derived by the Harvard group \cite{harvard} except that the
reactive term $\bm{\Omega} \times \bm{n}$ depends only on the spin
frequency $\bm{\Omega}$ and not on $\bm{A}$ and $\bm{\omega}$. If
$\gamma^{-1} = 0$, the director simply rotates like a rigid-body
axis perpendicular to the angular spin $\bm{\Omega}$.  When
$\gamma^{-1}$ is non zero, director motion decays to
$\bm{\Omega}\times\bm{n}$ in a time of order $K/(\gamma q^2)$
where $K$ is a Frank elastic constant and $q$ is the spatial
wavenumber of director distortions. Equations (\ref{0.4}) and
(\ref{0.5}) are the torque equations for spin angular momentum.
The spin frequency $\Omega_{\|}$ parallel to the director,
described by Eq.\ (\ref{0.5}), is a nonhydrodynamic variable that
decays in microscopic times to $\omega_{\|}$ plus nonlinear terms.
It cannot be ignored except in the hydrodynamic limit or in the
limit in which only the component of spin angular momentum
perpendicular to the director survives, i.e., when $\calI_{\|} =
0$,
which occurs in the limit of perfect nematic order in a system
composed of infinitely thin rods.
Equation (\ref{0.5}) contains a
nonlinear term proportional to $\Omega_{\|} \bm{\Omega}_{\perp}$
that has physical significance. As reviewed in Appendix\
\ref{appendix:rigid}, it causes the axis with the highest moment
of inertia to align along the direction of the angular momentum in
rigid body motion with conserved angular momentum but not energy.
The right hand sides of Eqs.\ (\ref{0.4}) and (\ref{0.5}) are the
time rate of change, $d \bm{l}/dt$, of the spin angular momentum
projected, respectively, along directions perpendicular and
parallel to $\bm{n}$.  In the absence of dissipation, $d\bm{l}/dt$
is simply the torque density $\bm{n}\times\bm{h}$ appearing in
Eq.\ (\ref{0.4}).  Rigid-body rotation in which $\bm{\Omega} =
\bm{\omega}$ and $\bm{A} = 0$ is a stationary state in which
$d\bm{l}/dt=0$.  If $\bm{\Omega}\neq\bm{\omega}$, dissipative
torques, given by the $\Gamma_{\perp}^{\Omega}$ and
$\Gamma_{||}^{\Omega}$ terms in Eq.\ (\ref{0.4}) and (\ref{0.5}),
drive $\bm{\Omega}$ towards $\bm{\omega}$.  Spin angular momentum
is also reoriented via the $\Gamma^A$ term in Eq.\ (\ref{0.4}) by
the symmetric strain rate $\bm{A}$ when it is nonzero. The stress
tensor contains a couple of terms not found in isotropic fluids.
The anti-symmetric parts of the stress tensor proportional to
$\Gamma_A$ and $\Gamma^{\Omega}_{ij}$ are dictated by the
requirement that the total spin and center-of-mass angular
momentum is conserved (see Appendix\ \ref{sec.app}). The
$\Gamma^A$ contribution to the symmetric part of the dissipative
srress tensor $\sigma_{ij}^{S\prime}$ is a consequence of an
Onsager relation.

When $\Gamma_{\|}^{\Omega}$, $\Gamma_{\perp}^{\Omega}$, and
$\Gamma^A$ are zero, spin angular momentum is conserved, and
additional diffusive dissipative terms proportional to $\nabla^2
\bm{\Omega}$ must be added to Eqs. (\ref{0.4}) and (\ref{0.5}) for
a complete description.  In this limit, Eq.\ (\ref{0.3}) along
with Eqs.\ (\ref{0.4}) and (\ref{0.5}) provide a hydrodynamical
description of rigid rotors on a rigid lattice with frictionless
bearings, which exhibit spin-wave excitations with a frequency
dispersion $\omega \sim q$ with $q$ the wavenumber
\cite{Chaikin-Lub}.

We can now consider under what conditions our equations reduce to
the original LE equations and how the hydrodynamical limit is
obtained. We begin with obtaining the LE equations. To approach
the LE limit, we use Eq.\ (\ref{0.3}) to replace $\bm{\Omega}
\times \bm{n}$ by $(d\bm{n}/d t) - \gamma^{-1} \bm{h}^T$, where
$\bm{h}^T = \bm h + \Delta \calI \Omega_{\|} \bm{\Omega}_{\perp}$.
This converts Eq.\ (\ref{0.4}) to
\begin{align}
&\calI_{\perp}\bm{n} \times \Big(\frac{d^{2}\bm{n}}{dt^{2}} - \frac{1}{\gamma }
\frac{d \bm{h}^T}{dt}\Big)+ \calI_{\|} \Omega_{\|} \frac{d\bm{n}}{d t} -
\frac{\Gamma_{\perp}^{\Omega} \Delta \calI}{\gamma}\Omega_{\|} \bm{n}
\times \bm{\Omega}
 \nonumber \\
& = \alpha \bm{n} \times (\bm{h} - \gamma _1 \bm{N} - \gamma_2
\bm{A} \bm{n}) ,
\label{eq:LE-p}
\end{align}
where we introduced
\begin{equation}
\bm{N} = \frac{d \bm{n}}{d t} - \bm{\omega} \times \bm{n}
\end{equation}
and
\begin{equation}
\alpha   = 1 + \frac{\Gamma^{\Omega}_{\perp}}{\gamma} ;
 \quad \frac{1}{\gamma_1 } = \frac{1}{\gamma } +
\frac{1}{\Gamma_{\perp}^{\Omega}};
 \quad \gamma_2 =  - \frac{\Gamma^A}{\Gamma_{\perp}^{\Omega}} \gamma_1
 \equiv  - \lambda \gamma_1 \enspace .
\label{eq:new_diss}
\end{equation}
Equation (\ref{eq:LE-p}) reduces identically to the original LE
equation\ \cite{LE} for the director with
left-hand side equal to
$\calI_{\perp} \bm{n} \times d^{2}\bm{n}/dt^{2}$,
$\alpha = 1$ and $\gamma_1 =
\Gamma_{\perp}^{\Omega}$, when $\gamma\rightarrow \infty$ and
$\calI_{\|} = 0$. The first condition, $\gamma\rightarrow\infty$
constrains $d\bm{n}/dt$ to be $\bm{\Omega} \times \bm{n}$.  The
second condition, $\calI_{\|} = 0$, is equivalent to there being
no rotational kinetic energy associated with $\Omega_{\|}$ and is
one that is tacitly assumed in the original LE approach for which
the spin kinetic energy density is $\calI_{\perp} (\bm{n} \times
\dot{\bm{n}})^2/2$.  We will show that $\calI_{\|}$ vanishes for
rigid rods when the Maier-Saupe order parameter $S$ equals one,
i.e., only when there is perfect order. If nematogens are modelled
by more complex rigid structures than thin rods, $I_{\|}$ would be
nonzero even for $S = 1$.
In the LE limit, Eq.\ (\ref{0.5}) for
$\Omega_{\|}$ implies that $\Omega_{\|} = \omega_{\|}$. Together
with $\bm{\Omega}_{\perp} - \bm{\omega}_{\perp} = \bm{n} \times
\bm{N}$, which follows from $d\bm{n} / dt = \bm{\Omega}\times \bm{n}$
and the definition for $\bm{N}$, the equations for the stress
tensor [Eqs. (\ref{0.7}) and (\ref{0.9})] assume exactly the form
of the LE stress tensor.
Thus to reiterate, the LE equations describe
a nematic liquid crystal in which the director is forced to follow
$\bm{\Omega}\times \bm{n}$ and the moment of inertial density
parallel to the director is zero.  Neither of these conditions
apply in general.

To obtain the hydrodynamic limit, we discard all terms that are
higher order in time and space derivatives than the dominant ones.
This means that we can ignore the $d \Omega_{\|}/dt$ and the
nonlinear terms in Eq.\ (\ref{0.5}) relative to
$\Omega_{\|}-\omega_{\|}$.  Thus to hydrodynamic order, we can set
$\Omega_{\|} = \omega_{\|}$.  This procedure effectively removes
$\Omega_{\|}$ from the problem.  Similarly, we can ignore the
$\Omega_{\|}$ term in the director equation [Eq.\ (\ref{0.3})] and
all of the terms on the
left-hand side of Eq.\ (\ref{eq:LE-p}).
The latter condition gives the familiar LE equation, $\bm{h} =
\gamma_1 \bm{N} + \gamma_2 (\bm{A} \bm{n})_{\perp}$,
for the director in
which the inertial term is neglected. To obtain the hydrodynamic
limit for the stress tensor, we use the hydrodynamic limit of
Eqs.\ (\ref{0.3}) and (\ref{eq:LE-p}) and the relations in Eq.\
(\ref{eq:new_diss}) to set
\begin{equation}
(\bm{\Omega} - \bm{\omega}) \times \bm{n} =
\frac{\gamma_1}{\Gamma_{\perp}^{\Omega}} \bm{N}
-\frac{\gamma_2}{\gamma} (\bm{A}\bm{n})_{\perp}
\end{equation}
and obtain
\begin{align}
\sigma_{ij}^{A \prime} & = \frac{1}{2}\gamma_1(n_i N_j - n_j N_i )
+ \frac{1}{2} \gamma_2 [n_i (\bm{A}\bm{n})_{\perp j} - n_j
(\bm{A}\bm{n})_{\perp i} ]\\
\sigma_{ij}^{S\prime} & = \alpha_{ijkl} A_{kl} + \frac{1}{2}
\gamma_2 (n_i N_j + n_j N_i)  \nonumber \\
& \qquad - \frac{1}{2} \frac{(\Gamma^A)^2}{\gamma + \Gamma_{\perp}^{\Omega}}
[n_i (\bm{A}\bm{n})_{\perp j} + n_j
(\bm{A}\bm{n})_{\perp i} ] .
\end{align}
This is precisely the LE stress tensor in the hydrodynamic limit.

An important consequence of this analysis that treats angular
momentum as an independent variable is that it demonstrates that
two distinct effects contribute to the viscosity $\gamma_1$: the
director damping measured by $\gamma$ and the rotational friction
measured by $\Gamma_{\perp}^{\Omega}$. $\gamma_1$ is the parallel
combination of $\gamma$ and $\Gamma_{\perp}^{\Omega}$
[see Eq.\ (\ref{eq:new_diss})].

In Ref.\ \cite{Stark2003} we derived dynamical equations for the
full nematic order parameter $\bm{Q}$, also called alignment
tensor, using the Poisson-bracket formalism without, however,
introducing the spin angular momentum density as a separate
dynamic variable. With the approach presented in the following, we
could also derive dynamic equations for $\bm{Q}$ and then by
projection on the uniaxial part of $\bm{Q}$ arrive at an
additional dynamic equation for the scalar order parameter. A
similar consideration following Eqs.
(\ref{eq:LE-p})-(\ref{eq:new_diss}) should then lead to the
extended LE equations of Ref.\ \cite{Ericksen91}, where a variable
$S$ is taken into account. The field $S$ is a nonhydrodynamic
variable that relaxes in microscopic times. It does not, however,
contribute to dissipative coefficients in the hydrodynamic limit
as the spin angular momentum, which also has a rapidly decaying
nonhydrodynamic component, does [see Eq.\ (\ref{eq:new_diss})]. We
will, therefore, not treat $S$ (or the biaxial part of $\bm{Q}$)
in what follows.

In the remainder of the article we give a detailed account of how
the set of equations discussed in this section were derived. We start
with a short review of the Poisson-bracket formalism.

\section{General formalism} \label{sec.form}

In this section we collect the important formulas of the
Poisson-bracket formalism. A more thorough explanation including
original references can be found in our previous article\ \cite{Stark2003}
and in\ \cite{Chaikin-Lub}.

We consider a systems whose microscopic dynamics is determined by
canonically conjugate variables $\bm{q}^{\alpha}$
and $\bm{\pi}^{\alpha}$ for each particle $\alpha$ and a
microscopic Hamiltonian ${\widehat {\cal H}}(\{\bm{q}^{\alpha}
\}, \{\bm{\pi}^{\alpha}\})$. Rotational degrees of freedom may be
included in the coordinates  $\bm{q}^{\alpha}$ and momenta
$\bm{\pi}^{\alpha}$. We are interested in the slow dynamics of
of a set of macroscopic field variables
$\Phi_{\mu}(\bm{x},t)$ ($\mu = 1,2, \ldots$)
obtained from microscopic fields $\widehat{\Phi}_{\mu}(\bm{x},
\{\bm{q}^{\alpha}\}, \{\bm{\pi}^{\alpha}\})$ by coarse-graining
over spatial fluctuations on the microscopic level;
$\Phi_{\mu}(\bm{x},t) = [ \, \widehat{\Phi}_{\mu}(\bm{x},
\{\bm{q}^{\alpha}\}, \{\bm{\pi}^{\alpha}\}) \, ]_{c} $, where
the symbol $[ \ldots ]_{c}$ specifies the coarse-grained averages.
The statistical mechanics of the macroscopic fields
$\Phi_{\mu}(\bm{x}, t)$ is determined by the coarse-grained
Hamiltonian ${\cal H}[\{\Phi_{\mu}\}]$.

Following the theory of
kinetic or stochastic equations, the macroscopic variables evolve
according to
\begin{equation}
\frac{\partial \Phi_{\mu}(\bm{x},t)}{\partial t} =
V_{\mu}(\bm{x}) - \Gamma_{\mu\nu} \,
\frac{\delta \mathcal{H}}{\delta \Phi_{\nu}(\bm{x})} \enspace,
\label{1.1}
\end{equation}
where we disregard any noise. The reactive term $V_{\mu}(\bm{x})$,
also called the non-dissipative or streaming velocity, is
expressed with the help of Poisson brackets as
\begin{equation}
V_{\mu}(\bm{x}) = - \int d^{3}x' \,
{\mathcal{P}_{\mu \nu}} ( \bm{x} , \bm{x}' ) \,
\frac{\delta \mathcal{H}}{\delta \Phi_{\nu}(\bm{x}')} \enspace,
\label{1.2}
\end{equation}
where Einstein's convention on repeated indices is understood and
\begin{equation}
{\mathcal{P}_{\mu \nu}}( \bm{x} , \bm{x}')
= \{ \Phi_{\mu}(\bm{x}), \Phi_{\nu}(\bm{x}')\}
=-{\mathcal{P}_{\nu \mu}} ( \bm{x}',\bm{x} )
\label{1.3}
\end{equation}
denotes the Poisson bracket of the coarse-grained variables.
It is defined as the coarse-grained average of the microscopic
Poisson bracket:
\begin{equation}
\{\Phi_{\mu} ( \bm{x} ) , \Phi_{\nu} ( \bm{x}' )\} =
[ \{\widehat{\Phi}_{\mu} ( \bm{x} ) ,
\widehat{\Phi}_{\nu} (\bm{x}') \} ]_{c} \enspace,
\label{1.4}
\end{equation}
where\ \cite{Goldstein1983}
\begin{eqnarray}
\lefteqn{
\{\widehat{\Phi}_{\mu}(\bm{x}),\widehat{\Phi}_{\nu}(\bm{x}')\}
 =  }
\nonumber \\
 & & \sum_{\alpha i}
\frac{\partial \widehat{\Phi}_{\mu}(\bm{x})}{\partial
\pi^{\alpha}_{i}}
\frac{\partial \widehat{\Phi}_{\nu}(\bm{x}')}{\partial
q^{\alpha}_{i}} -
\frac{\partial \widehat{\Phi}_{\mu}(\bm{x})}{
     \partial q^{\alpha}_{i}}
\frac{\partial \widehat{\Phi}_{\nu}(\bm{x}')}{
     \partial \pi^{\alpha}_{i}} \enspace.
\label{1.5}
\end{eqnarray}

Since we only employ a restricted number of macroscopic variables,
all the ``neglected'' microscopic degrees of freedom give rise to
the dissipative term in the kinetic equation\ (\ref{1.1}) that is
proportional to the generalized force $\delta {\cal
H}/\delta{\Phi_{\nu}(\bm{x})}$, which together with
$\Phi_{\nu}(\bm{x})$ forms a pair of conjugate variables. The
dissipative tensor $\Gamma_{\mu\nu}$ may depend on the fields
$\Phi_{\mu}$ and it may also contain terms proportional to
$-\bm{\nabla}^{2}$. It is determined by three principles. First,
the dissipative contributions to the equation for  $\partial
\Phi_{\mu}/\partial t$ must have the same sign under time reversal
as $\Phi_{\mu}$ (and thus the opposite sign to that of $\partial
\Phi_{\mu}/ \partial t$). Second, $\Gamma_{\mu\nu}$ has to
reflect the local point group symmmetry of the dynamical system,
and third, it has to be a symmetric tensor at zero magnetic field
to obey the Onsager principle\ \cite{Groot1951}. In the following,
the last point will be important in identifying the proper
dissipative terms in the momentum balance.

\section{Poisson brackets for nematic liquid crystals} \label{sec.poisson}

\subsection{Model molecule and dynamic variables} \label{sec.nemlc}

We model our system as a fluid of uniaxial rigid rods of length
$a$ and mass $m$. We describe the position of molecule $\alpha$ by
its center-of-mass coordinate $\bm{x}^{\alpha}$ and its
orientation by the unit vector $\hat{\bm{\nu}}^{\alpha}$. The
center-of-mass momentum is $\bm{p}^{\alpha}= m \bm{v}^{\alpha} = m
\dot{\bm{x}}^{\alpha}$ where dot means total time derivative. With
the help of the molecular order-parameter tensor $Q^{\alpha}$ with
components
\begin{equation}
\textstyle
Q_{ij}^{\alpha} = \hat{\nu}_{i}^{\alpha} \hat{\nu}_{j}^{\alpha}
- \frac{1}{3} \delta_{ij} \enspace,
\label{3.1}
\end{equation}
the molecular moment-of-inertia tensor (relative to the center of mass)
reads
\begin{equation}
\textstyle I_{ij}^{\alpha} = I_{||} \hat{\nu}_i^{\alpha}
\hat{\nu}_j^{\alpha} + I_{\perp} (\delta_{ij} -
\hat{\nu}_{i}^{\alpha} \hat{\nu}_{j}^{\alpha}) = \Delta I
Q_{ij}^{\alpha}+(\frac{2}{3}I_{\perp} + \frac{1}{3}
I_{||})\delta_{ij}
\label{3.2}
\end{equation}
where $I_{\perp}$ and $I_{||} < I_{\perp}$ are the moments of
inertia for rotations about axes perpendicular and parallel to
$\hat{\bm{\nu}}^{\alpha}$, respectively, and $\Delta I= I_{||} -
I_{\perp}$. Note that $I_{||} = 0$, and $\Delta I = - I_{\perp}$
in the limit of infinitely thin rods. The tensor $\bm{I}^{\alpha}$
relates the spin angular momentum of a molecule,
$\bm{l}^{\alpha}$, to its angular velocity $\bm{\Omega}^{\alpha}$:
\begin{equation}
l^{\alpha}_{i} = I_{ij}^{\alpha} \Omega_{j}^{\alpha} \enspace.
\label{3.3}
\end{equation}
Note that $\bm{l}^{\alpha}$ is always perpendicular to
$\hat{\bm{\nu}}^{\alpha}$ as it should be for an infinitly thin rod.

The microscopic Poisson bracket in Eq.\ (\ref{1.5}) consists of a
spatial and an angular part. The spatial contribution arises
from the coordinate $\bm{x}^{\alpha}$ and its conjugate momentum
$\bm{p}^{\alpha}$, which fulfill the canonical Poisson
bracket\ \cite{Goldstein1983}:
\begin{equation}
\{p_{i}^{\alpha} , x_{j}^{\beta}\} = \delta^{\alpha\beta} \delta_{ij}
\enspace,
\label{3.4}
\end{equation}
where $\delta^{\alpha\beta}$ and $ \delta_{ij}$ are Kronecker symbols.
However, the unit vector $\hat{\bm{\nu}}^{\alpha}$ and the
angular momentum $\bm{l}^{\alpha}$ are not canonically conjugate
to each other since their Poisson bracket\ \cite{Goldstein1983}
\begin{equation}
\{l_{i}^{\alpha} , \hat{\nu}_{j}^{\beta} \} = - \delta^{\alpha\beta}
\varepsilon_{ijk} \hat{\nu}_{k}^{\beta}\enspace,
\label{3.5}
\end{equation}
does not have the canonical form ($\varepsilon_{ijk}$
denotes the Levi-Civita symbol). We could now introduce appropriate
pairs of conjugate angular coordinates and momenta via a microscopic
Legendre function. Instead, we follow an alternative route. It turns
out that, in the following, Eq.\ (\ref{3.5}) and the additional formula
\begin{equation}
\{l_{i}^{\alpha} , l_{j}^{\beta} \} = - \delta^{\alpha\beta}
\varepsilon_{ijk} l_{k}^{\beta}
\label{3.6}
\end{equation}
are sufficient for calculating the angular part of Poisson
bracket\ (\ref{1.5}). All other Poisson brackets, in particular
the ones between the spatial and angular variables, are zero.
The Poisson brackets in Eqs. (\ref{3.5}) and (\ref{3.6}) are a
consequence of the fact that $l_i^{\alpha}$ is the generator of
rotations about the molecular center of mass
\cite{Dzyaloshinskii1980}.  They can, or course, be derived from a
Hamiltonian formalism for rigid rods in which the rotational
kinetic energy depends on two Euler angles for the case of
infinitely thin rods ($I_{||} = 0$) and three Euler angles for the
more general case.

We are now ready to define the relevant microscopic field variables
and their coarse-grained counterparts. The conventional microscopic
definition of the density of mass and center-of-mass momentum are:
\begin{eqnarray}
\widehat{\rho}(\bm{x}) & = & m \sum_{\alpha}
\delta(\bm{x}-\bm{x}^{\alpha})
\label{3.7} \\
\widehat{\bm{g}}(\bm{x}) & = &
\sum_{\alpha} \bm{p}^{\alpha} \delta(\bm{x}-\bm{x}^{\alpha})
\enspace,
\label{3.8}
\end{eqnarray}
which, after coarse graining, result in the macroscopic variables
$\rho(\bm{x}) = [\widehat{\rho}(\bm{x}) ]_{c}$ and
$\bm{g}(\bm{x}) = [\widehat{\bm{g}}(\bm{x}) ]_{c} \equiv
\rho(\bm{x}) \bm{v}(\bm{x})$. The last term defines the
macroscopic velocity field $\bm{v}(\bm{x})$. In a similar manner
we introduce the macroscopic field $\bm{Q}(\bm{x})$ of the
nematic tensorial order parameter\
\cite{deGennes1993,Saupe1964,deGennes1969,Lubensky1970},
also called the alignment tensor\ \cite{Hess1975}:
\begin{equation}
\frac{\rho(\bm{x})}{m} \bm{Q}(\bm{x}) =
\left[
\sum_{\alpha} \bm{Q}^{\alpha} \delta(\bm{x}-\bm{x}^{\alpha})
\right]_{c}
\label{3.9}
\end{equation}
using $\bm{Q}^{\alpha}$ from Eq.\ (\ref{3.1}). The factor
$\rho(\bm{x})/m$ is introduced to make $\bm{Q}(\bm{x})$ unitless.
With the microscopic definition for the density of the
moment-of-inertia tensor,
\begin{align}
\widehat{\calI}_{ij}(\bm{x}) & =  \sum_{\alpha} I_{ij}^{\alpha}
\delta(\bm{x}-\bm{x}^{\alpha}) \nonumber \\
& = \sum_{\alpha} [\Delta I Q_{ij}^{\alpha} +({\textstyle
\frac{2}{3}}I_{\perp} +{\textstyle \frac{1}{3}} I_{||})
\delta_{ij}] \delta(\bm{x}-\bm{x}^{\alpha}) \enspace,
\label{3.10}
\end{align}
we obtain the coarse-grained moment-of-inertia density
\begin{equation}
\calI_{ij}(\bm{x}) = \frac{\rho(\bm{x})}{m} \, [ \Delta I
Q_{ij}(\bm{x}) +({\textstyle \frac{2}{3}}I_{\perp} +{\textstyle
\frac{1}{3}} I_{||}) \delta_{ij}] \enspace.
\label{3.11}
\end{equation}
Finally, the microscopic field of the density of spin angular momentum is
\begin{equation}
\widehat{\bm{l}}(\bm{x}) = \sum_{\alpha} \bm{l}^{\alpha}
\delta(\bm{x}-\bm{x}^{\alpha}) \enspace.
\label{3.12}
\end{equation}
Its associated coarse-grained variable is $\bm{l}(\bm{x}) =
[\widehat{\bm{l}}(\bm{x})]_{c} \equiv
\bm{\calI}(\bm{x}) \bm{\Omega}(\bm{x})$, where the last expression
defines the macroscopic field $\bm{\Omega}(\bm{x})$ of angular velocity
in full analogy to $\bm{v}(\bm{x})$.

We are interested in the dynamics of the nematic phase where the
orientational order is uniaxial. The alignment tensor therefore
assumes the form
\begin{equation}
\textstyle
Q_{ij}(\bm{x}) = S [n_{i}(\bm{x}) n_{j}(\bm{x})
- \frac{1}{3} \delta_{ij} ] \enspace,
\label{3.13}
\end{equation}
where, on average, the molecules point along the director
$\bm{n}(\bm{x})$. The Maier-Saupe order parameter $S$ is constant
in the nematic phase. With the uniaxial $\bm{Q}(\bm{x})$ of
Eq.\ (\ref{3.13}), the moment-of-inertia density $\bm{\calI}(\bm{x})$
of Eq.\ (\ref{3.11}) becomes
\begin{equation}
\calI_{ij} = \calI_{\|} n_{i} n_{j} + \calI_{\perp} (\delta_{ij}
   - n_{i} n_{j})
\label{3.13a}
\end{equation}
with
\begin{eqnarray}
\calI_{\|} & = &  \frac{\rho}{m} \left( \frac{2}{3} I_{\perp}
+\frac{1}{3} I_{||} + \frac{2}{3} \Delta I S \right) \nonumber \\
\calI_{\perp} & = &  \frac{\rho}{m} \left( \frac{2}{3} I_{\perp}
+\frac{1}{3} I_{||} - \frac{1}{3} \Delta I S \right)\enspace.
\label{3.13b}
\end{eqnarray}
Its anisotropy is quantified by
\begin{equation}
\Delta \calI = \calI_{\|} - \calI_{\perp} = \Delta I
\frac{\rho}{m} S \enspace .
\label{3.13c}
\end{equation}
Note, as indicated, in Sec.\ \ref{sec.review}, that $\calI_{\|} =
0$ in the limit of infinitely thin rods ($I_{||} = 0$) and perfect
nematic order ($S=1$).

Thus, our set of dynamic variables is $\{\rho,\bm{n},\bm{g},\bm{l} \}$
for which we have to determine all possible Poisson brackets.

\subsection{Poisson Brackets} \label{subsec.nemlc.PB}
The calculation of the Poisson brackets is straightforward. In
addition to the comments about the angular variables in the
previous section and the antisymmetry relation expressed in
Eq.\ (\ref{1.3}), we use properties of the $\delta$ function summarized as
\begin{subequations}
\label{3.14}
\begin{eqnarray}
\delta(\bm{x} - \bm{x}') & = & \delta(\bm{x}' - \bm{x}) \\
f(\bm{x}) \delta(\bm{x} - \bm{x}') & = &
f(\bm{x}') \delta(\bm{x} - \bm{x}') \\
\nabla_{i} \delta(\bm{x} - \bm{x}') & = &
- \nabla'_{i} \delta(\bm{x} - \bm{x}') \enspace,
\end{eqnarray}
\end{subequations}
where $\nabla_{i} = \partial / \partial x_{i}$,
$\nabla'_{i} = \partial / \partial x'_{i}$, and $f(\bm{x})$ is an
arbitrary function including the $\delta$ function itself.

In the following, we list all the non-zero Poisson brackets which
determine the non-dissipative velocities of our dynamic variables.
The dynamics of the center-of-mass density $\rho(\bm{x})$ is
provided by
\begin{equation}
\{ \rho(\bm{x}),g_{i}(\bm{x}')\} =
\nabla_{i} \delta(\bm{x} - \bm{x}') \rho(\bm{x}')
\enspace.
\label{3.15}
\end{equation}

To derive the Poisson brackets of the director, we first calculate
the Poisson brackets of the alignment tensor. According to the
definition\ (\ref{3.9}), we only have a microscopic expression for
$\rho(\bm{x}) \bm{Q}(\bm{x})$ but not for $\bm{Q}(\bm{x})$ alone.
An analogous, however more complicated, situation occurred in our
previous article\ \cite{Stark2003}. To calculate, {\em e.g.},
$\{Q_{ij}(\bm{x}), g_{k}(\bm{x}') \}$, we apply the Product rule
for Poisson brackets to $\{ \rho(\bm{x}) Q_{ij}(\bm{x}),
g_{k}(\bm{x}') \}$ and arrive at
\begin{eqnarray}
\{Q_{ij}(\bm{x}), g_{k}(\bm{x}') \} & = &
\frac{1}{\rho(\bm{x})}
\{ \rho(\bm{x}) Q_{ij}(\bm{x}), g_{k}(\bm{x}') \} \nonumber\\
& & \qquad  - \frac{Q_{ij}(\bm{x})}{\rho(\bm{x})}
\{\rho(\bm{x}), g_{k}(\bm{x}') \} \enspace.
\label{3.16}
\end{eqnarray}
The first term on the right-hand side and the second term, already known
from Eq.\ (\ref{3.15}), then combine to yield
\begin{equation}
\{Q_{ij}(\bm{x}), g_{k}(\bm{x}') \} = \nabla_{k}
\delta(\bm{x}-\bm{x}') Q_{ij}(\bm{x}') \enspace.
\label{3.17}
\end{equation}
In the same manner, we calculate
\begin{equation}
\{Q_{ij}(\bm{x}), l_{k}(\bm{x}') \} =
- [\varepsilon_{ijk}Q_{il}(\bm{x}) + \varepsilon_{ikl}Q_{jl}(\bm{x})]
\delta(\bm{x}-\bm{x}') \enspace,
\label{3.18}
\end{equation}
where we have used the product rule and Eq.\ (\ref{3.5}) to evaluate the
microscopic Poisson bracket $\{Q_{ij}^{\alpha}, l_{k}^{\beta}\}$ and
the fact that $\{\rho(\bm{x}),l_{k}(\bm{x}')\} = 0$. The Poisson
brackets for the director now follow by projection from the uniaxial
representation (\ref{3.13}) of the alignment tensor
(see Ref.\ \cite{Stark2003} for details):
\begin{equation}
\{n_i(\bm{x}) , g_j (\bm{x}')\} = \frac{1}{S} \delta_{ik}^T
\{ Q_{kl}(\bm{x}), g_j (\bm{x}')\} n_l(\bm{x}) \enspace ,
\label{3.19}
\end{equation}
where
\begin{equation}
\delta^{T}_{ij} = \delta_{ij} - n_{i} n_{j}
\label{3.20}
\end{equation}
is the projector on the space perpendicular to $\bm{n}(\bm{x})$.
The same formula is valid with $g_j (\bm{x}')$ replaced by
$l_j (\bm{x}')$ so that Eqs.\ (\ref{3.17})-(\ref{3.20}) finally give
\begin{subequations}
\label{3.21}
\begin{eqnarray}
\{n_i(\bm{x}) , g_j (\bm{x}')\} & = &
[\nabla_{j} n_{i}(\bm{x})] \delta(\bm{x}-\bm{x}')
\label{3.21a} \\
\{n_i(\bm{x}) , l_j (\bm{x}')\} & = & -\varepsilon_{ijk}
n_{k}(\bm{x}) \delta(\bm{x}-\bm{x}') \enspace.
\label{3.21b}
\end{eqnarray}
\end{subequations}
Note that the Poisson brackets of $Q_{ij}(\bm{x})$ and $n_{i}(\bm{x})$ with
$g_{k}(\bm{x}')$ are much simpler in this formulation with spin
angular momentum than they are in the alternative one\
\cite{Stark2003,Forster1974}
in which there is no spin angular momentum and $\bm{g}$ is
the total rather than the center-of-mass momentum density. In
particular, the director-momentum bracket $\lambda_{ijk} \nabla_{k}
\delta(\bm{x}-\bm{x}')$ that plays such an important role in the
latter formulation is not present in the current one.

The reactive velocity of the translational momentum follows from
\begin{subequations}
\label{3.22}
\begin{eqnarray}
\{g_{i}(\bm{x}),\rho(\bm{x}')\} & = &
\rho(\bm{x}) \nabla_{i} \delta(\bm{x} - \bm{x}')
\label{3.22a} \\
\{g_{i}(\bm{x}), n_{j}(\bm{x}')\} & = &
-[\nabla_{i} n_{j}(\bm{x})] \delta(\bm{x} - \bm{x}')
\label{3.22b} \\
\{g_{i}(\bm{x}), g_{j}(\bm{x}')\} & = &
-\nabla'_{i}[\delta(\bm{x}-\bm{x}')g_{j}(\bm{x}')]
\nonumber \\
& & + \nabla_{j}\delta(\bm{x}-\bm{x}') g_{i}(\bm{x}')
\label{3.22c} \\
\{g_{i}(\bm{x}), l_{j}(\bm{x}')\} & = & l_{j}(\bm{x})
\nabla_{i} \delta(\bm{x}-\bm{x}') \enspace.
\label{3.22d}
\end{eqnarray}
\end{subequations}
Eqs.\ (\ref{3.22a}) and (\ref{3.22b}) are related to Eqs.\ (\ref{3.15})
and (\ref{3.21a}) by the antisymmetry relation of the Poisson
brackets whereas Eqs.\ (\ref{3.22c}) and (\ref{3.22d}) are readily
calculated. Again the missing term in the momentum-director bracket
compared to Ref.\ \cite{Stark2003} is compensated by the additional Poisson
bracket\ (\ref{3.22d}).

Finally, the non-dissipative dynamics of the angular-momentum density
is governed by
\begin{subequations}
\label{3.23}
\begin{eqnarray}
\{l_{i}(\bm{x}),n_{j}(\bm{x}')\} & = & -\varepsilon_{ijk}
n_{k}(\bm{x}) \delta(\bm{x} - \bm{x}')
\label{3.23a} \\
\{l_{i}(\bm{x}), g_{j}(\bm{x}')\} & = &
l_{i}(\bm{x}') \nabla_{j} \delta(\bm{x} - \bm{x}')
\label{3.23b} \\
\{l_{i}(\bm{x}), l_{j}(\bm{x}')\} & = & -\varepsilon_{ijk}
l_{k}(\bm{x}') \delta(\bm{x}-\bm{x}') \enspace.
\label{3.23c}
\end{eqnarray}
\end{subequations}

\section{Nematodynamics with Spin Angular Momentum} \label{sec.nemato}

Following the systematic structure of the theory outlined in
Sec.\ \ref{sec.form}, we now derive the full set of equations as
presented and discussed in Sec.\ \ref{sec.review}. We first
calculate the reactive and dissipative velocities needed to formulate
the dynamic eqations for the set of dynamic variables
$\{\rho,\bm{n},\bm{g},\bm{l}\}$ and then introduce the spin angular
velocity $\bm{\Omega}$.

\subsection{Non-dissipative velocities}
To calculate the non-dissipative velocities from Eq.\ (\ref{1.2}), we
need the Hamiltonian
\begin{eqnarray}
{\cal H} & = &\int \Big[ \frac{\bm{g}^{2}(\bm{x})}{2\rho(\bm{x})}
+ \frac{1}{2} l_{i}(\bm{x}) \calI^{-1}_{ij}(\bm{x})
l_{j}(\bm{x}) \Big] d^{3}x \nonumber \\
& &  \qquad \quad + F[\rho(\bm{x}),\bm{n}(\bm{x})] \enspace.
\label{4.1}
\end{eqnarray}
It consists of a kinetic part, subdivided into a translational and
rotational term, and a free energy
$F[\rho(\bm{x}),\bm{n}(\bm{x})] = \int
f(\rho,\bm{n},\bm{\nabla}\bm{n}) d^{3}x$, which is Frank's free
energy plus a term depending only on $\rho$.
In the following, we will need derivatives of the inverse of the
moment-of-inertia density such as
$\partial \calI^{-1}_{ij} / \partial y$
where $y$ stands for $\rho$ or $\bm{n}$. Taking the
derivative of $ \calI_{ij} \calI_{jk}^{-1} = \delta_{ik}$ with respect to $y$,
we find
\begin{equation}
\frac{\partial}{\partial y} \calI^{-1}_{ij} = - \calI^{-1}_{ik}
\Big(\frac{\partial}{\partial y}\calI_{kl} \Big)  \calI^{-1}_{lj}
\enspace.
\label{4.3}
\end{equation}

The non-dissipative velocity for the density is simply
$V^{\rho} = - \bm{\nabla} \cdot \bm{g}(\bm{x})$ which immediately gives
the mass-conservation law. For the director, we use the Poisson
brackets (\ref{3.21}) and the fact that
$\delta {\cal H}/\delta l_{j}(\bm{x}') = \calI^{-1}_{jk}l_{k} =
\Omega_{j}(\bm{x}')$ and arrive at
\begin{equation}
V^{\bm{n}}_{i} = -\bm{v}(\bm{x}) \cdot \bm{\nabla}
 n_{i}(\bm{x}) + \varepsilon_{ijk} \Omega_{j}(\bm{x})
n_{k}(\bm{x}) \enspace .
\label{4.4}
\end{equation}
The first term on the right-hand side is the convective derivative
of $\bm{n}$. The second term introduces a reactive coupling to the
angular velocity $\bm{\Omega}(\bm{x})$.  In the purely
hydroydnamical model \cite{harvard,Stark2003}, this term is
replaced by $\lambda_{ijk} \nabla_j v_k$ coupling $\partial n_i/
\partial t$ to the symmetric and anti-symmetric parts of the
deformation-rate tensor $\nabla_i v_j$

The non-disspative velocity $\bm{V}^{\bm{g}}$ is calculated with
the help of the Poisson brackets\ (\ref{3.22}). Applied to the
rotational part of the kinetic energy in ${\cal H}$, i.e.,
$\frac{1}{2} l_{i}(\bm{x}) \calI^{-1}_{ij}(\bm{x}) l_{j}(\bm{x})$,
they produce a contribution $\bm{V}^{\bm{g}}_{\mathrm{rot}}$ to
the non-dissipative velocity $\bm{V}^{\bm{g}}$ whose terms add up
to zero. Specifically, we find
\begin{eqnarray}
V^{\bm{g}}_{\mathrm{rot},i} & = & -\rho \nabla_{i}
\Big(\frac{1}{2}l_{k}l_{l} \frac{\partial
\calI_{kl}^{-1}}{\partial \rho} \Big)
\nonumber\\
& &  + (\nabla_{i}n_{j}) \Big(\frac{1}{2} l_{k}l_{l}\frac{\partial
\calI^{-1}_{kl}}{\partial n_{j}}\Big) -l_{j} \nabla_{i}\Omega_{j}
 \enspace,
\label{4.5}
\end{eqnarray}
which we rewrite as
\begin{eqnarray}
V^{\bm{g}}_{\mathrm{rot},i} & = & \nabla_{i} \Big(\frac{1}{2}
\calI_{kl}^{-1}l_{k}l_{l}\Big)
\nonumber \\
& & -\nabla_{i}\Big(\rho \frac{1}{2}l_{k}l_{l}
\frac{\partial \calI_{kl}^{-1}}{\partial \rho} \Big)
-\nabla_{i}(l_{j}\Omega_{j}) \enspace.
\label{4.6}
\end{eqnarray}
Using Eq.\ (\ref{4.3}) to evaluate the derivative with respect to
$\rho$ and the fact that $\calI_{kl}$ is linear in $\rho$
[see Eqs.\ (\ref{3.13a}) and (\ref{3.13b})], one shows immediately
that the three terms add up to zero, i.~e.,
$\bm{V}^{\bm{g}}_{\mathrm{rot}} = 0$.
All the other contributions to
$\bm{V}^{\bm{g}}$ can be written in a compact form,
\begin{equation}
V_{i}^{\bm{g}} =
-\nabla_{j} \Big[\frac{g_{i}(\bm{x}) g_{j}(\bm{x})}{\rho(\bm{x})}\Big]
-\nabla_{i}p + \nabla_{j} \sigma^{E}_{ij} \enspace,
\label{4.7}
\end{equation}
as explained in Ref.\ \cite{Stark2003}. The first term on the right-hand side
introduces the momentum flux tensor, the second term contains the
pressure $p$, and $\bm{\sigma}^{E}$ is the Ericksen stress tensor:
\begin{equation}
\sigma^{E}_{ij} =
- \frac{\partial f}{\partial \nabla_{j}n_{k}} \nabla_{i}n_{k} \enspace.
\label{4.8}
\end{equation}
Note that compared to Ref.\ \cite{Stark2003} a term that contains the
molecular field $\delta F/\delta \bm{n}$ is completely missing.

Finally the Poisson brackets (\ref{3.23}) give the reactive velocity
of the angular-momentum density:
\begin{eqnarray}
V_{i}^{\bm{l}} & = & \varepsilon_{ijk} n_{k}(\bm{x})
\Big(\frac{1}{2}l_{r}l_{s}\frac{\partial \calI_{rs}^{-1}}{\partial n_{j}}
+ \frac{\delta F}{\delta n_{j}(\bm{x})} \Big) \nonumber\\
 & & -\nabla_{j}[l_{i}(\bm{x})v_{j}(\bm{x})] + \varepsilon_{ijk}
\Omega_{j}(\bm{x}) l_{k}(\bm{x}) \enspace.
\label{4.9}
\end{eqnarray}
Again, one can show using Eqs.\ (\ref{3.13a}) and (\ref{4.3}) that the first
and the fourth term on the right-hand side resulting from the
rotational kinetic energy cancel each other so that we obtain
\begin{equation}
V_{i}^{\bm{l}} = -\nabla_{j}[l_{i}(\bm{x})v_{j}(\bm{x})]
+ \varepsilon_{ijk} n_{k}(\bm{x})
\frac{\delta F}{\delta n_{j}(\bm{x})} \enspace.
\label{4.10}
\end{equation}
The first term introduces the angular momentum flux tensor in full
analogy to the linear momentum and the second term is a coupling to
the molecular field.

\subsection{Dissipative velocities and final equations}

\subsubsection{Center-of-mass density}
For the conserved center-of-mass density, dissipative velocities are
not allowed, and the mass-conservation law follows:
\begin{equation}
\frac{\partial \rho}{\partial t} = - \bm{\nabla} \cdot \bm{g} \enspace.
\label{4.11}
\end{equation}

\subsubsection{Director}
The time derivative $\partial \bm{n} / \partial t$ couples
dissipatively only to forces conjugate to fields, $\bm{n}$ and
$\rho$ with the same sign under time reversal as $\bm{n}$. A
dissipative term proportional to $ \bm{n} \delta {\cal H} / \delta
\rho$, which has the correct sign under time reversal, cannot
occur because it is always perpendicular to $\partial
\bm{n}/\partial t$ \cite{Stark2003}. A second dissipative term
introduces a coupling to $\delta {\cal H} / \delta n_{i}$ with a
dissipative tensor $\delta^{T}_{ij}/\gamma$, where the projector
$\delta^{T}_{ij}$ defined in Eq.\ (\ref{3.20}) ensures that
$\partial \bm{n} /
\partial t$ is perpendicular to $\bm{n}$ and $\gamma$ is a
rotational viscosity. Together with Eq.\ (\ref{4.4}), the dynamic
equation for the director then reads
\begin{equation}
\frac{\partial n_{i}}{\partial t} = -\bm{v} \cdot \bm{\nabla} n_{i}
+ \varepsilon_{ijk} \Omega_{j} n_{k} -\frac{1}{\gamma}
\delta_{ij}^{T} \Big(\frac{1}{2} l_{k}l_{l}
\frac{\partial \calI_{kl}^{-1}}{\partial n_{j}} +
\frac{\delta F}{\delta n_{j}} \Big) \enspace.
\label{4.12}
\end{equation}
With the definition of the components of the molecular field,
$h_{i} = -\delta_{ij}^{T} \delta F/\delta n_{j}$, and the expression
\begin{equation}
\delta_{ij}^{T} \frac{1}{2} l_{k}l_{l}
\frac{\partial \calI_{kl}^{-1}}{\partial n_{j}} =
-(\calI_{\|} - \calI_{\perp}) \Omega_{\|} \bm{\Omega}_{\perp} \enspace,
\end{equation}
which we derive with the help of Eqs. (\ref{3.13a}) and (\ref{4.3})
and $l_i = \calI_{ij} \Omega_{j}$, we finally arrive at the director
equation as presented in Eq.\ (\ref{0.3}) in Sec.\ \ref{sec.review}.

\subsubsection{Spin angular momentum density}

The fields $\delta {\cal H}/\delta l_{i} = \Omega_{i}$ and $\delta
{\cal H}/\delta g_{i} = v_{i}$  have the same sign under time
reversal as $\bm{l}$ and can contribute terms of the equation for
$\partial \bm{l}/\partial t$. To determine the form of these
terms, it is important to realize that $\bm{l}$ and $\bm{\Omega}$
are pseudo vectors that do not change sign under space inversion
but that the momentum density, $\bm{g}$, and the velocity,
$\bm{v}$, are vectors that do change sign under space inversion.
Thus a term directly proportional to $ v_{i}$ in the equation for
$\partial l_i/\partial t$ is prohibited but one proportional to
$\Omega_{i}$ is not. Pseudo vectors that are even under $\bm{n}
\rightarrow - \bm{n}$  can be constructed from the spatial
derivatives of $\bm{v}$ and the director. Thus we look for
dissipative terms containing these pseudovector combinations of
$\nabla_{i}v_{j}$. The first pseudovector is $\omega_{i} =
\varepsilon_{ijk}\nabla_{j}v_{k}$/2. Together with the angular
velocity, it gives rise to the dissipative term
$-\Gamma^{\Omega}_{ij} (\Omega_{j} - \omega_{j})$ whose form is
dictated by the requirement that during a uniform rotation of the
whole sample ($\bm{\Omega} = \bm{\omega}$) no energy is
dissipated. A second pseudo vector of the velocity, which
preserves the $\bm{n} \rightarrow -\bm{n}$ symmetry of the nematic
phase, is $\frac{1}{2} (\varepsilon_{ijl}n_{l}n_{k} +
\varepsilon_{ikl}n_{l}n_{j}) A_{jk}$, where $A_{jk} = (\nabla_{j}
v_{k} + \nabla_{k} v_{j})/2$ stands for the symmetrized velocity
gradient. Furthermore, the third-rank tensor in front of $A_{jk}$
is symmetric in $j$ and $k$ which is important for the next
paragraph where we use the Onsager principle to find the
dissipative velocities for the momentum density. Introducing the
dissipative torque
\begin{equation}
\tau'_{i} =  -\Gamma^{\Omega}_{ij} (\Omega_{j} - \omega_{j}) -
\frac{\Gamma^{A}}{2}
(\varepsilon_{ijl}n_{l}n_{k} + \varepsilon_{ikl}n_{l}n_{j}) A_{jk} \enspace,
\label{4.13}
\end{equation}
where
\begin{equation}
\Gamma^{\Omega}_{ij} = \Gamma^{\Omega}_{\|} n_{i}n_{j} +
\Gamma^{\Omega}_{\perp} (\delta_{ij}-n_{i}n_{j})
\label{4.14}
\end{equation}
obeys the uniaxial symmetry of the nematic phase, and combining it
with Eq.\ (\ref{4.10}), we arrive at the formula describing the
dynamics of $\bm{l}$:
\begin{equation}
\frac{\partial l_{i}}{\partial t} =  -\nabla_{j} (l_{i} v_{j}) +
\varepsilon_{ijk} n_{k} \frac{\delta F}{\delta n_{j}} +
\tau'_{i} \enspace.
\label{4.15}
\end{equation}
To replace $\bm{l}$ by the angular velocity, we write
\begin{equation}
\bm{l} = \calI_{\|} \Omega_{\|} \bm{n} + \calI_{\perp} \bm{\Omega_{\perp}}
\enspace .
\label{4.15a}
\end{equation}
The time derivative of $\bm{l}$ involves $\partial \calI_{\alpha}
/\partial t$, where $\alpha$ represents $\|$ or $\perp$. We find
\begin{equation}
\frac{\partial \calI_{\alpha}}{\partial t} =
\frac{\partial \calI_{\alpha}}{\partial \rho}
\frac{\partial \rho}{\partial t} = -\nabla_{i} (\calI_{\alpha} v_{i})
\enspace,
\label{4.15b}
\end{equation}
where we used the fact that $\calI_{\alpha}$ linearly depends on $\rho$
[see Eq.\ (\ref{3.13b})] and where we also employed the
mass-conservation law $\partial \rho/\partial t = -\nabla_{i}(\rho v_{i})$.
With Eq.\ (\ref{4.15b}) and the definition of the total time
derivative, $d /d t = \partial /\partial t + v_{i} \nabla_{i}$,
it is straightforward to show that
\begin{equation}
\frac{\partial l_{i}}{\partial t} + \nabla_{j} (l_{i} v_{j}) =
\calI_{\perp} \frac{d \bm{\Omega}_{\perp}}{d t} +
\Omega_{\|} \frac{d \bm{n}}{d t} + \bm{n} \frac{d \Omega_{\|}}{d t}
\enspace .
\label{4.15c}
\end{equation}
We introduce this term into the balance equation (\ref{4.15}) for
$\bm{l}$ and project it on $\bm{n}$ and the plane perpendicular to $\bm{n}$
to finally arrive at the respective Eqs.\ (\ref{0.5}) and (\ref{0.4})
in our review section\ \ref{sec.review}.
In deriving the last two terms of Eq.\ (\ref{0.5}), we also used
$\bm{n} \cdot d\Omega_{\perp}/d t = - (d \bm{n}/d t) \cdot \Omega_{\perp}$
and replaced $d \bm{n}/d t$ by Eq.\ (\ref{0.3})

\subsubsection{Center-of-mass momentum density}
The dissipative term of the momentum balance is determined by the
viscous stress tensor $\bm{\sigma}'$ which couples again to $
\Omega_{i}$ and $\nabla_{i}v_{j}$ as in the previous paragraph.
The form of the dissipative part of the stress tensor is subject
to restrictions. First of all, because the total angular momentum
(spin plus center-of-mass) is conserved, the antisymmetric part
$\bm{\sigma}^{A \prime}$ of the viscous stress tensor and the dissipative
torque $\bm{\tau}'$ of Eq.\ (\ref{4.13}) are related (see Appendix\
\ref{sec.app}):
\begin{eqnarray}
\sigma_{ij}^{A \prime} & = &
- \frac{1}{2} \varepsilon_{ijk} \tau_{k}' \nonumber\\
 & = & \frac{1}{2} \varepsilon_{ijk} \Gamma^{\Omega}_{kl}
(\Omega_{l} - \omega_{l} ) \nonumber \\
 & & + \frac{1}{2}\Gamma^{A}
(n_{j}A_{in}n_{n} - n_{i}A_{jn}n_{n}) \enspace.
\label{4.16}
\end{eqnarray}
To construct the symmetric part $\bm{\sigma}^{S \prime}$ of the viscous
stress tensor, we use Onsager's principle. It says that the
dissipative fluxes $\bm{\tau}'$, $\bm{\sigma}^{S \prime}$ are coupled to
the generalized forces $\bm{\Omega}-\bm{\omega}$, $\bm{A}$
by a symmetric, dissipative tensor. In symbolic notation this means
\begin{equation}
\left(\begin{array}{c} -\bm{\tau}' \\ \bm{\sigma}^{S \prime} \end{array}\right)
=
\left(
\begin{array}{cc}
\bm{\Gamma}^{\Omega} & \Gamma^{A}\bm{\varepsilon n n} \\
\Gamma^{A}(\bm{\varepsilon n n})^{t} & \bm{\alpha}
\end{array}
\right) \,
\left(\begin{array}{c} \bm{\Omega}-\bm{\omega} \\
      \bm{A} \end{array}\right) \enspace,
\label{4.17}
\end{equation}
where the superscript $t$ in $(\bm{\varepsilon n n})^{t}$ stands for the
appropriately transposed third-rank tensor $\bm{\varepsilon n n}$
of Eq.\ (\ref{4.13}). The first line of the tensor equation
reproduces the dissipative torque\ (\ref{4.13}),
the second line gives
\begin{equation}
\sigma_{ij}^{S \prime} = \alpha_{ijkl} A_{kl} + \frac{1}{2} \Gamma^{A}
(\varepsilon_{ilk}n_{j}n_{l} + \varepsilon_{jlk} n_{i}n_{l})
(\Omega_{k}-\omega_{k}) \enspace,
\label{4.18}
\end{equation}
where the viscosity tensor $\bm{\alpha}$ has the usual form
required by the uniaxial symmetry of our medium
(see, e.~g., Ref.\ \cite{Stark2003}):
\begin{align}
& \alpha_{ijkl} = \alpha_{1} n_{i} n_{j} n_{k} n_{l}
+ \frac{\alpha_{4}}{2} (\delta_{ik} \delta_{jl} + \delta_{il}
\delta_{jk})
\nonumber\\
& +\frac{\alpha_{5} + \alpha_{6}}{4}
(n_{i}n_{k} \delta_{jl} + n_{j}n_{k} \delta_{il} + n_{i}n_{l}
\delta_{jk}
 + n_{j}n_{l} \delta_{ik}) \nonumber \\
& + \zeta_{1} \delta_{ij} \delta_{kl} + \zeta_{2} (\delta_{ij}
n_{k}n_{l} + n_{i} n_{j} \delta_{kl}) \enspace.
\label{4.19}
\end{align}
Adding up reactive and dissipative terms, the momentum balance
finally reads
\begin{equation}
\frac{\partial g_{i}}{\partial t} = -\nabla_{j}
\Big(\frac{g_{i}  g_{j}}{\rho }\Big)
-\nabla_{i}p + \nabla_{j} ( \sigma^{E}_{ij} + \sigma_{ij}^{S \prime}
+ \sigma_{ij}^{A \prime}) \enspace.
\label{4.20}
\end{equation}

The complete set of equations of nematodynamics including the spin
angular momentum is reproduced in Eqs.\ (\ref{0.1}) to
(\ref{0.9}). The Leslie-Ericksen and hydroydnamic limits of these
equations are derived in Sec.\ \ref{sec.review}.

\acknowledgments
H.S. acknowledges financial support from the Deutsche
Forschungsgemeinschaft under grant No. Sta 352/5-1 and through the
International Graduate College ``Soft Matter''.
T.C.L. was supported by the US National Science Foundation under
grant No. DMR04-05187.

\appendix

\section{\label{appendix:rigid} Nonlinearities in Rigid-Body Motion}

Our equation for the director reduces to that for the anisotropy
axis $\bm{n}$ for a single uniaxial rigid body if $\bm{n}(\bm{x})$
is independent of $\bm{x}$ and $\bm{h} = 0$.  If
$\Gamma^{\Omega}_{ij} = 0$ and $\Gamma^A = 0$, spin angular
momentum is conserved [See Eq.\ (\ref{4.15})], i.e., $d\bm{L}/dt =
d/dt(\int d^3x \bm{l}) = 0$, and $\Omega_{||} =
L_{||}/\tilde{I}_{||}$ and $\bm{\Omega}_{\perp} = \bm{L}_{\perp}/
\tilde{I}_{\perp}$ where $\tilde{I}_{||,\perp} = \int d^3 x
\calI_{||,\perp}$. The equation of motion for the anisotropy axis
is then
\begin{equation}
\frac{d \bm{n}}{dt} = \bm{\Omega} \times \bm{n} + \frac{\Delta
\calI }{\gamma}\Omega_{||} \bm{\Omega}_{\perp} .
\label{eq:AA1}
\end{equation}
This equation has a dissipative term implying that energy is not
conserved even though spin angular momentum is. Since $\bm{L}$ is
a constant, we can choose it to be a vector $L \bm{e}_z$ of fixed
length pointing along the space fixed unit vector $\bm{e}_z$ along
the $z$-axis. When $\bm{n} = (\sin \theta \cos \phi, \sin \theta
\sin\phi, \cos \theta)$ is expressed in polar coordinates relative
to the $z$ axis, Eq.\ (\ref{eq:AA1}) reduces to the equations
\begin{eqnarray}
\frac{d \phi}{dt} & = & \frac{L}{\tilde{I}_{\perp}} \label{eq:AA2}\\
\frac{d \theta}{dt} & = & - \frac{1}{2 \gamma} \frac{\Delta
\calI}{\tilde{I}_{||} \tilde{I}_{\perp}} L^2 \sin 2 \theta .
\label{eq:AA3}
\end{eqnarray}
The equation for $\theta$ is easily solved subject to the boundary
condition that $\theta ( t = 0 ) = \theta_0$:
\begin{equation}
\tan \theta(t) = \tan \theta_0 \exp\left[- \frac{1}{4 \gamma}
\frac{\Delta \calI}{\tilde{I}_{||} \tilde{I}_{\perp}} L^2 t
\right].
\end{equation}
Thus, if $\calI_{||} > \calI_\perp$, $\theta(t) \rightarrow 0$ if
$0\leq\theta_0<\pi/2$ or $\theta(t) \rightarrow \pi$ if
$\pi/2<\theta_0\leq\pi$ as $t \rightarrow \infty$.  This means
that $\bm{n}$ will align or anti-align with the angular momentum
direction and that the angular momentum comes entirely from
spinning parallel to the anisotropy axis with kinetic energy
$L^2/(2 \tilde{I}_{||})$. If $\calI_{||}< \calI{\perp}$,
$\theta(t) \rightarrow \pi/2$ for $0<\theta_0<\pi$. In this case,
$\bm{n}$ lies in the $xy$-plane and rotates according to Eq.\
(\ref{eq:AA2}), and the kinetic energy is $L^2/(2
\tilde{I}_{\perp})$.  Thus, when angular momentum is conserved but
energy is not, the rigid body will evolve toward the state with
the lowest kinetic energy consistent with the constraint of fixed
angular momentum.

\section{Torques and Stress Tensor} \label{sec.app}
Here we shortly demonstrate that the antisymmetric part of the
stress tensor is equivalent to a torque acting on the intrinsic
angular momentum. The total angular momentum of a system with
volume V is given by
\begin{equation}
\bm{L} = \int_{V} (\rho \bm{x} \times \bm{v} + \bm{l}) d^{3}x
\enspace.
\label{A1}
\end{equation}
In case of zero body forces and torques (which might originate from
applied magnetic and electric fields or gravitation), only surface
forces can change the total angular momentum. Per definition they are
described by the stress tensor $\bm{\sigma}$ so that
\begin{equation}
\frac{d\bm{L}}{dt} = \int_{V} \bm{x} \times \bm{\sigma} d\bm{f}
\enspace,
\label{A2}
\end{equation}
where $\partial V$ means surface of $V$. Applying Gauss's theorem to the
right-hand side results in
\begin{equation}
\frac{d\bm{L}}{dt} = \int_{\partial V} (\bm{x} \times
{\mathrm div}\bm{\sigma} + \bm{\tau} ) d^{3}x
\label{A3}
\end{equation}
where we introduced the torque
\begin{equation}
\tau_{i} = - \varepsilon_{ijk} \sigma_{jk} \enspace.
\label{A4}
\end{equation}
With
\begin{equation}
\frac{d\bm{L}}{dt} = \int_{V}
\Big(\rho \bm{x} \times \frac{d\bm{v}}{dt} + \frac{d\bm{l}}{dt} \Big)
\label{A5}
\end{equation}
and the momentum balance in differential form,
\begin{equation}
\rho \frac{d\bm{v}}{dt} = \mathrm{div} \bm{\sigma} \enspace,
\label{A6}
\end{equation}
we obtain from Eq.\ (\ref{A3})
\begin{equation}
\frac{d\bm{l}}{dt} = \bm{\tau} \enspace,
\end{equation}
where $\bm{\tau}$ is the torque acting on $\bm{l}$.


\begin{thebibliography}{10}

\bibitem{LE}
J.~L. {Ericksen}, Arch.\ Ratl.\ Mech.\ Anal. \textbf{4}, 231 (1960);
Trans.\ Soc.\ Rheol. \textbf{5}, 23 (1961);
F.~M. {Leslie}, Quart.\ J.~Mech.\ Appl.\ Math. \textbf{19}, 357 (1966);
Arch.\ Ratl.\ Mech.\ Anal. \textbf{28}, 265 (1968).

\bibitem{Carlsson99}
T. Carlsson and F.~M. Leslie, Liq.\ Cryst.\ \textbf{26}, 1267 (1999).

\bibitem{harvard}
D.~{Forster}, T.~C.~{Lubensky}, P.~C.~{Martin}, J.~{Swift}, and
P.~S.~{Pershan}, Phys.\ Rev.\ Lett. \textbf{26}, 1016 (1971).

\bibitem{deGennes1993} P.~G. de~{Gennes} and J.~{Prost}, \emph{The Physics
    of Liquid Crystals}, 2nd edition (Oxford Science Publications,
    Oxford, 1993).

\bibitem{Martin}
P.~C. Martin, O. Parodi, and P.~S. Pershan, Phys.\ Rev.\ A
\textbf{6}, 2401 (1972).

\bibitem{Zwanzig1961}
R. Zwanzig, Phys.\ Rev. \textbf{124}, 983 (1961);
R. Zwanzig in \emph{Statistical Mechanics: New Concepts, New Problems,
  New Applications}, ed. by S.~A. Rice, K.~F. Freed, and J.~C. Light,
  The University of Chikago Press, Chikago (1972).

\bibitem{Kawasaki1970}
K. Kawasaki, Ann.\ Phys.\ (N.Y.) \textbf{61}, 1 (1970);
K. Kawasaki in \emph{Critical Phenomena}, ed. by M.~S. Green,
Academic, New York (1971).

\bibitem{Mori1973}
H. Mori and H. Fujisaka, Progr.\ Theor.\ Phys.\ \textbf{49}, 764 (1973);
H. Mori, H. Fujisaka, and H. Shigematsu, Progr.\ Theor.\ Phys.\
\textbf{51}, 109 (1973).

\bibitem{Ma1975}
S. Ma and G.~F. Mazenko, Phys.\ Rev.~B \textbf{11}, 4077 (1975).

\bibitem{Chaikin-Lub}
P.~{Chaikin} and T.~C. {Lubensky}, \emph{Principles of Condensed Matter
  Physics}, Cambridge University Press, Cambridge (1995).

\bibitem{Stark2003}
H. Stark and T.~C. Lubensky, Phys.\ Rev. E\ \textbf{67}, 061709 (2003).

\bibitem{Ericksen91}
J.~L. Ericksen, Arch. Rational Mech. Anal.\ \textbf{113}, 97 (1991).

\bibitem{Dzyaloshinskii1980}
I.~E. Dzyaloshinskii and G.~E. Volovick, Ann.\ Phys. \textbf{125}, 67 (1980).

\bibitem{Grmela}
M. Grmela, Phys.\ Lett. A \textbf{130}, 81 (1988);
A.~N. Beris and B.~J. Edwards, J.\ Rheol. \textbf{34}, 55 (1990).

\bibitem{Oettinger97}
H.~C. \"Ottinger and M. Grmela, Phys.\ Rev.\ E\ \textbf{56}, 6620 (1997);
Phys.\ Rev.\ E\ \textbf{56}, 6633 (1997).


\bibitem{Lubensky1970}
T.~C. Lubensky, Phys.\ Rev.\ A \textbf{2}, 2497 (1970).

\bibitem{Goldstein1983}
H. Goldstein, \emph{Klassische Mechanik}, 7th edition (Ak\-a\-de\-mi\-sche
Verlagsgesellschaft, Wiesbaden, 1983).

 \bibitem{Groot1951}
S.~R. de Groot, \emph{Thermodynamics of Irreversible Processes},
North-Holland Publishing Company, Amsterdam (1951).

\bibitem{Saupe1964}
A. Saupe, Z.~Naturforsch.\ \textbf{19a}, 161 (1964).

\bibitem{deGennes1969}
P.~G. de~{Gennes}, Phys.\ Lett.\ \textbf{30A}, 454 (1969).

\bibitem{Hess1975}
S. Hess, Z.~Naturforsch.\ \textbf{30a}, 728 (1975).

\bibitem{Forster1974}
D. Forster, Phys.\ Rev.\ Lett. \textbf{32}, 1161 (1974).
\end{thebibliography}
\end{document}